\documentclass[fleqn,10pt]{wlscirep}
\usepackage[utf8]{inputenc}
\usepackage[T1]{fontenc}
\usepackage{bm}
\usepackage{comment}
\usepackage{caption}
\usepackage{amsmath}
\usepackage{color}
\usepackage{lineno}

\usepackage[acronym]{glossaries}
\usepackage{geometry}
\usepackage{hyperref}
\title{RingSim- An Agent-based Approach for Modelling Mesoscopic Magnetic Nanowire Networks}
\author{Ian T Vidamour$^{1,2}$, Guru Venkat$^{1}$, Charles Swindells$^{1}$, David Griffin$^{3}$, Paul W Fry$^{4}$, Richard M Rowan-Robinson$^{1}$, Alexander Welbourne$^{1}$, Francesco Maccherozzi$^{5}$, Sarnjeet S Dhesi$^{5}$, Susan Stepney$^{3}$, Dan A Allwood$^{1}$, and Thomas J Hayward$^{1}$\\
    \small School of Chemical, Materials, and Biological Engineering, University of Sheffield$^{1}$ \\
    \small School of Computer Science, University of Sheffield$^{2}$ \\
    \small Department of Computer Science, University of York$^{3}$ \\
    \small Centre for Nanoscience and Technology, University of Sheffield$^{4}$ \\
    \small Diamond Light Source Ltd., Science and Technology Facilities Council UK$^{5}$
}
\begin{abstract}
We describe "RingSim", a phenomenological agent-based model that allows numerical simulation of magnetic nanowire networks with areas of hundreds of micrometers squared for durations of hundreds of seconds; a practical impossibility for general-purpose micromagnetic simulation tools. In RingSim, domain walls (DWs) are instanced as mobile agents which respond to external magnetic fields, and their stochastic interactions with pinning sites and other DWs are described via simple phenomenological rules. We first present a detailed description of the model and its algorithmic implementation for simulating the behaviours of arrays of interconnected ring-shaped nanowires, which have previously been proposed as hardware platforms for unconventional computing applications. The model is then validated against a series of experimental measurements of an array's static and dynamic responses to rotating magnetic fields. The robust agreement between the modeled and experimental data demonstrates that agent-based modelling is a powerful tool for exploring mesoscale magnetic devices, enabling time scales and device sizes that are inaccessible to more conventional magnetic simulation techniques. \\
\end{abstract}
\begin{document}
\flushbottom
\maketitle
\section{Introduction}
The creation of models of system behaviour is critical to the development of emerging technologies, since they allow for rapid evaluation of device designs without the need for manufacturing samples or performing experimental measurements. For devices based around magnetic materials, the processes that underpin the device's response to external inputs often have simple analytical models. For example, devices such as spin-torque nano-oscillators \cite{torrejon_neuromorphic_2017, kanao_reservoir_2019-1}, domain wall (DW) oscillators \cite{PhysRevB.93.144410,ababei_neuromorphic_2021}, and super-paramagnetic ensembles \cite{welbourne_voltage-controlled_2021} have one-dimensional numerical descriptions that allow the state of the magnetic substrate to be approximated to predict device performance quickly with low computation cost.\\\\
For devices that are not characterised well by simple analytical models, or when more detail of the underlying magnetic state of the system is required, the typical approach is to use general-purpose micromagnetic simulation packages such as MuMax3 \cite{vansteenkiste_design_2014} or OOMMF \cite{donahue1999oommf}. These platforms approximate the spin structure of the magnetic materials into cells on the order of a few nanometers in size, and model the evolution of spins via the Landau-Lifschitz-Gilbert equation \cite{landau1992theory, gilbert1955lagrangian}. While these simulation packages provide a high level of detail on the magnetic response of a device, they are associated with a high computational overhead. For example, in simulations of a Skyrmion confined in a nanodisk of 80nm diameter, 1nm thickness, using a mesh size of 1nm$^{3}$, simulating 50ms of dynamic response takes on the order of 40 minutes \cite{chen_forecasting_2022}, a simulation duration 48,000 times longer than the physical processes being simulated, despite the vast acceleration of these simulations possible with modern hardware \cite{leliaert2019tomorrow}. \\ \\
While magnetic systems with fast dynamics such as spin-torque nano-oscillators require simulations at the nanosecond/microsecond timescale, other systems with dynamics governed by thermal processes, such as the nanowire network presented here, may respond on much slower timescales and hence need orders of magnitude longer simulation durations. When coupled with the relatively larger size of these systems, micromagnetic simulation approaches becomes practically impossible. Despite these challenges, the huge design space of networks of interacting magnetic elements such as artificial spin-ice systems \cite{skjaervo2020advances}, or arrays of interconnected magnetic nanorings \cite{dawidek_dynamically-driven_2021}, makes exploration via simulations very attractive for prototyping new computational platforms. These systems also exhibit emergent behaviours, where interactions between elements in the system lead to global behaviours that cannot be described by the action of individual elements alone. While these complex dynamics provide technical difficulty for simulation, they can be exploited for neuromorphic computing purposes \cite{jensen_reservoir_2017, jensen_reservoir_2020,gartside_reconfigurable_2022,hon2021numerical, stenning_adaptive_2022, vidamour_quantifying_2022, vidamour2023reconfigurable, venkat2024exploring}. Therefore, it is clear that alternative approaches for modelling such systems must be taken. \\ \\
Agent-based models provide a potential solution to these problems. These models describe the evolution of complex, multi-element systems in terms of the interactions between individual agents, as well as external environmental parameters \cite{5429318}. The agents are often instanced into the model with distributed parameters, and commonly exhibit stochastic behaviours. Interactions between agents are programmed phenomenologically, with the outcomes of interactions determined by a set of predefined rules that aim to encapsulate the behaviours of the system being simulated. These types of models are especially harmonious with systems that exhibit emergent behaviours, and have been used extensively in modelling complex dynamic systems such as flocking birds \cite{stonedahl2011finding}, or even structures within the brain \cite{joyce2012complexity, zhang2009simulating}.\\ \\
In this paper, we introduce an agent-based model, RingSim, which allows rapid simulation of networks of thousands of interconnected magnetic nanorings over timescales on the order of seconds. The model simulates the stochastic pinning, depinning and propagation of DWs at local pinning sites that occur at the junctions between rings in the arrays. DWs are instanced as agents that interact with these pinning sites, as  well as other DWs and externally applied magnetic fields. We fit model parameters describing ring junction properties and DW-DW interactions, and show how the model produces excellent with experimental data. Not only is the model capable of capturing the system's global response to external inputs both statically and dynamically, but also provides information on the microstate of each ring, visually showing similar local agglomeration of domains and relative populations of individual ring states as observed experimentally.
\section{Magnetic Nanoring Array Dynamics}
RingSim describes the emergent response of interconnected magnetic nanoring arrays (Figure 1(a)), experimentally detailed in previous works \cite{dawidek_dynamically-driven_2021,vidamour_quantifying_2022,vidamour2023reconfigurable}. Conceptually, the response of the system to rotating magnetic fields can be described via the transitions between metastable domain configuration for each of the rings, mediated by stochastic pinning events. Figure 1(b) shows the three basic ring configurations: a 'vortex' state containing a single domain and no DWs an 'onion' state containing two equally sized domains and a pair of DWs at opposite ends of the ring, and a 'three-quarter' state, featuring two differently sized domains, with DWs situated at angles of 90 degrees from one another.\\ \\
\begin{figure}[ht!]
\includegraphics[width=150mm]{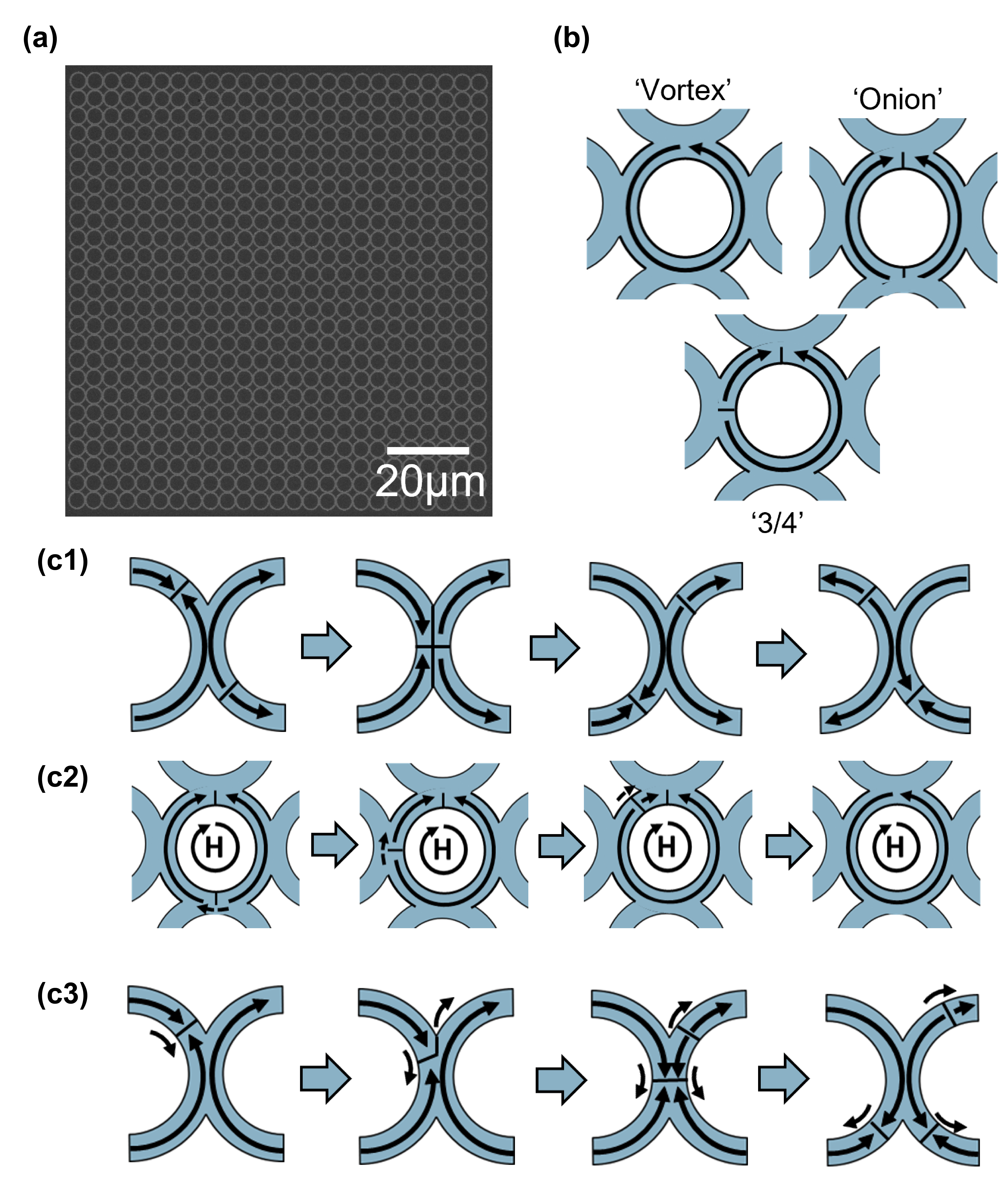}
\caption{(a)- Scanning electron micrograph showing a 25x25 magnetic nanoring array. (b)- Schematic diagrams showing the three metastable domain states in an individual ring: Vortex configuration, Onion configuration, and Three-Quarter configuration. Arrows show direction of magnetisation, and lines normal to the circumference of the rings reflect the position of domain walls. (c)- Schematics showing the outcomes of different stochastic propagation events: (c1)-Coherent propagation of DWs with sufficiently strong magnetic fields, (c2)-Stochastic propagation of DWs within a ring, with the upper domain wall becoming locally pinned before being annihilated by the propagating lower domain wall, and (c3)-The renucleation process when an itinerant domain wall causes local magnetic reversal, injecting a pair of domain walls into an adjacent ring. Arrows show magnetisation direction, with lines reflecting domain walls. Dotted arrows included to show propagation direction of domain walls, and large blue arrows represent the progression of time.}
\end{figure}
The interconnected nature of the ring arrays leads to local domain wall pinning sites at the junctions between neighbouring rings. The change in geometry presented by the junction creates an energy barrier, similar to an 'anti-notch' \cite{chang_magnetic_2011, sandweg_direct_2008}, which tends to locally pin itinerant domain walls. When driven with sufficiently high amplitude rotating fields, the domain walls are able to overcome this energy barrier, and coherently propagate with the rotating field, shown in Figure 1(c1). Under smaller driving fields, domain walls can become locally pinned at these junctions, with a finite probability of depinning via thermal activation. These stochastic processes can lead to the domain walls within a single rings propagating differently depending upon the outcomes of pinning events. This in turn may then lead to domain wall collision and subsequent annihilation, changing the ring state from an onion/three-quarter state to a vortex state, shown in Figure 1(c2).\\ \\
Domain walls may also be reintroduced into the system when propagating domain walls lead to magnetic reversal across a junction. To avoid causing magnetic frustration when the magnetisation of a junction is reversed, a pair of domain walls are nucleated in the ring adjacent to the propagating domain wall, shown in Figure 1(c3). Between these mechanisms for domain wall annihilation and renucleation, a dynamic equilibrium is created between the rate of domain wall loss and gain, depending upon the relative probabilities of domain wall pinning/depinning for a given applied field.\\ \\ 
At very low or very high applied fields, the pinning/propagation events are effectively deterministic, leading to few collisions and hence the array exists as mainly onion states with a few three quarter states. For intermediate applied fields, the stochastic movement of domain walls leaves the array in a mixture of states from all three configurations depending upon the relative rates of collision and renucleation. \\ \\
\section{Modelling Stochastic Pinning Events} The array behaviour can be approximated via simulation of domain state of each ring, determined by the outcomes of the pinning events. To achieve this, we use empirically verified relationships to calculate expected probabilities of domain walls in the system propagating beyond pinning sites. In magnetic materials, thermal energy introduces stochastic domain wall motion via the random fluctuation of individual magnetic moments which assist reversal processes. This results in a finite expected timescale for a reversal event to occur, depending upon the size of the associated energy barrier and the temperature of the system. Empirically, the Arrhenius-Néel relationship calculates the characteristic timescale of reversal via equation $(1)$: \\ \\$(1)\indent \tau_r = \tau_0^{\frac{\Delta E}{k_B T}}$ \\ \\ \indent where $\tau_r$ represents the expected reversal timescale, $\tau_0$ the reciprocal of attempt frequency $f_0$ associated with the material (here $f_0$ $\approx{}$1GHz for $\mathrm{Ni}_{80}\mathrm{Fe}_{20}$ \cite{lau2007common}), $\Delta E$ the magnitude of the effective energy barrier, $k_B$ the Boltzmann constant, and $T$ the temperature of the system. This relationship has been experimentally verified for the reversal of single magnetic domains, with excellent agreement \cite{wernsdorfer_experimental_1997}. \\ \\
As well as the temperature of the system, external magnetic fields also influence the outcome of pinning events by modulating the magnitude of the effective energy barrier. Previous work has shown that this modulation is dependent upon the component of applied field acting tangentially to the ring at the position of the domain wall, with the domain walls having lowest Zeeman energy when aligned with the field vector.\cite{negoita_controlling_2012, negoita_domain_2013}. The rotating magnetic fields used to drive the ring arrays means that this transverse component, $H_\mathrm{transverse}$ is dependent upon the magnitude of the applied field, $H_\mathrm{applied}$, and the angle between the applied field and the domain wall,  $\theta_{\mathrm{lag}}$,  shown schematically in Figure 2(a) and described mathematically via: \\ \\
$(2) \indent H_\mathrm{transverse} = H_\mathrm{applied} *sin(\theta_\mathrm{lag})$ \\ \\
\begin{figure}[ht!]
    \centering
    \includegraphics[width=\textwidth]{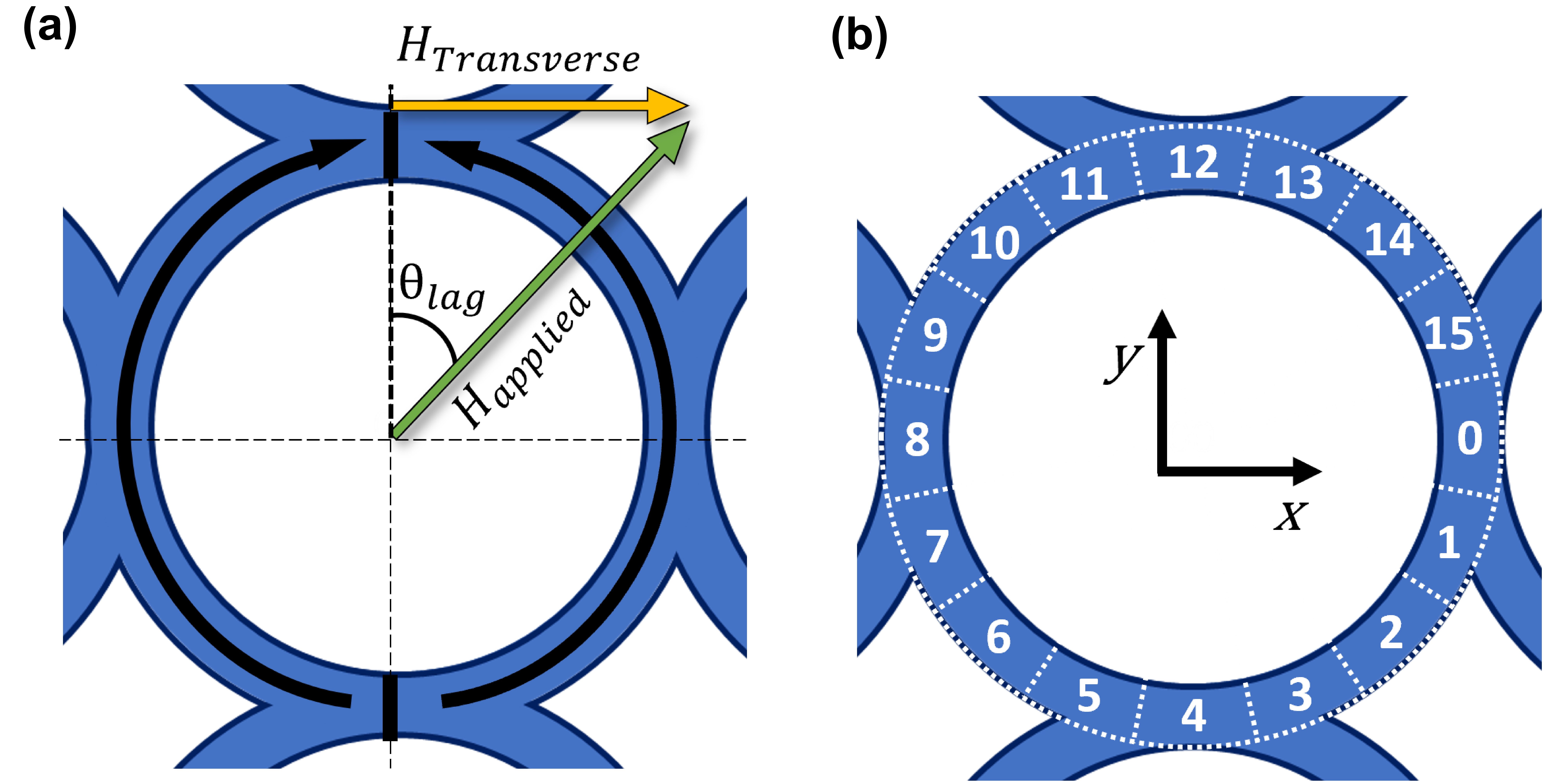}
    \caption{(a)- Schematic diagram showing the calculation of $H_\mathrm{transverse}$ from an applied field $H_\mathrm{applied}$ and the angular lag between the domain wall location and the direction of applied field, $\theta_\mathrm{lag}$. (b)- Schematic diagram representing the cardinal X and Y axes directions, as well as the corresponding junction indices at each of the intersections between the rings for $N_\mathrm{segment} = 16$.}
    \label{fig:fig2l}
\end{figure}
The relationship between the transverse field, the magnitude of the modulated energy barrier, $\Delta E$, and the zero-field barrier, $E_0$, is given via the phenomenological Sharrock equation \cite{sharrock_time_1994}: \\ \\
$(3) \indent \Delta E = E_0 (1 - \frac{H_\mathrm{transverse}}{H_\mathrm{sw}^0})^\alpha$ \\ \\
\indent where $H_\mathrm{sw}^0$ represents the zero-Kelvin switching field of the magnetic element, and $\alpha$ is a geometrical constant,  (here taken to be 3/2 \cite{wegrowe_magnetic_1997}).\\ \\
It has been observed in previous works that the presence of either one or two domain walls across a junction leads to different domain wall structures \cite{dawidek_dynamically-driven_2021}, each with different energetic properties, and hence different depinning behaviours. Here, we assume that $E_0$ and $H_\mathrm{sw}^0$ for the single and double domain wall case are related as follows: \\ \\
$(4) \indent H_\mathrm{2DW}^{0} = k * H_\mathrm{1DW}^{0}$ \\ \\
$(5) \indent E_\mathrm{2DW}^{0} = k * E_\mathrm{1DW}^{0}$ \\ \\
\indent where $H_\mathrm{1DW}^{0}$ and $H_\mathrm{2DW}^{0}$ represent the switching fields for one and two domain wall cases respectively, $E_\mathrm{1DW}^0$ and $E_\mathrm{2DW}^0$ the equivalents for initial energy barriers, and $k$ a fixed scaling parameter, the fitting of which is discussed in section 5. \\ \\
In order to approximate the varying $H_\mathrm{transverse}$ (and consequently $E_0$) as the applied field rotates, the field is discretised into a series of angular steps, each held for a duration of $t_\mathrm{step} = \frac{N_\mathrm{steps}}{2\pi f}$ seconds, where $N_\mathrm{steps}$ represents the number of discretisation steps per rotation, and $f$ represents the rotational frequency of the applied field. This allows the approximated depinning probability for a domain wall in a given discretisation step, $P_\mathrm{depin}$, to be calculated from equations $(1-5)$ via equation $(6)$: \\ \\
$(6) \indent P_\mathrm{depin} = 1 - e^{-\frac{t_\mathrm{step}}{\tau_r}}$ \\ \\

\section{Phenomenological Modelling of Magnetic Nanoring Arrays}
Phenomenological descriptions of domain walls and their interactions allow a simple method for programming experimentally observed propagation, annihilation, and nucleation behaviours. Parameters describing properties of rings, junctions, and numbers of elements have labels R, J, and N respectively. \\\\
The rings within the model are represented by vectors of length $N_\mathrm{segment}$, where each entry to the vector represents a ring segment of arc length $2\pi/N_\mathrm{segment}$ radians. The indices of this vector each represent a position within the ring, rotating clockwise from the positive x direction, shown in Figure 2(b). Here, a value of 16 for $N_\mathrm{segment}$ was selected to provide a good trade-off between approximating a smooth rotation of field and matching the fourfold symmetry of the array, whilst keeping the number of simulation steps low.\\ \\
The domain state of the nanoring array, $R_\mathrm{DW}$, is expressed as a matrix of dimensions $[N_\mathrm{segment} \times (N_{r})^2]$, where $N_{r}$ represents the number of rings in each row of the square array. Domain walls are instanced into the array by labelling an index in each ring vector with either a +1 or a -1, reflecting head-to-head and tail-to-tail domain walls respectively. Since many of the key behaviours of the ring array are determined by the junctions between the rings in the array and the interactions that arise at them, three separate vectors of length $2N_r(N_r - 1)$ are created which record the properties of every junction in the network: $J_{DW}$, which tracks the number of domain walls occupying each junction, $J_E$, which reflects the magnitude of the energy barrier $E_{0}$ presented by each junction, and $J_H$, which reflects the zero-kelvin switching field $H_{sw}^0$ for each junction,  as described in the previous section. \\ \\
The state of the simulated nanoring array is initialised by instancing a head-to-head domain wall in every ring in $R_\mathrm{DW}$ at index $\frac{N_\mathrm{segment}}{4}$, and a tail-to-tail domain wall in every index $\frac{3\times N_\mathrm{segment}}{4}$, corresponding to the positive/negative y direction respectively, and emulating the saturated state of all onion state rings aligned in the positive y direction. The magnetisation state of the array is then generated from the position and variety of all domain walls in the system. Firstly, an additional array, $R_\mathrm{dir}$, of identical shape to $R_\mathrm{DW}$ is generated. This array records whether the magnetisation runs clockwise (+1) or anticlockwise (-1) over each segment, and is marked zero in the locations of domain walls. From this direction array, the net magnetisation of the array may be calculated in terms of components in the x and y axes ($M_x$ and $M_y$ respectively) via:\\ \\
$(7) \indent M_x = \sum_{i=1}^{N_r^2}\sum_{s=1}^{N_\mathrm{segment}} \mathrm{sin}(\frac{2\pi s}{N_\mathrm{segment}})R_\mathrm{dir}^{i,s}$ \\ \\
$(8) \indent M_y = \sum_{i=1}^{N_r^2}\sum_{s=1}^{N_\mathrm{segment}} \mathrm{cos}(\frac{2\pi s}{N_\mathrm{segment}})R_\mathrm{dir}^{i,s} $ \\ \\
\indent This gives the magnetisation of the array in arbitrary units, which is then normalised against the magnetisation in the saturated state, determined as the value of $M_y$ in the initialised array.\\ \\
\begin{figure}[ht!]
    \centering
    \includegraphics[width=\textwidth]{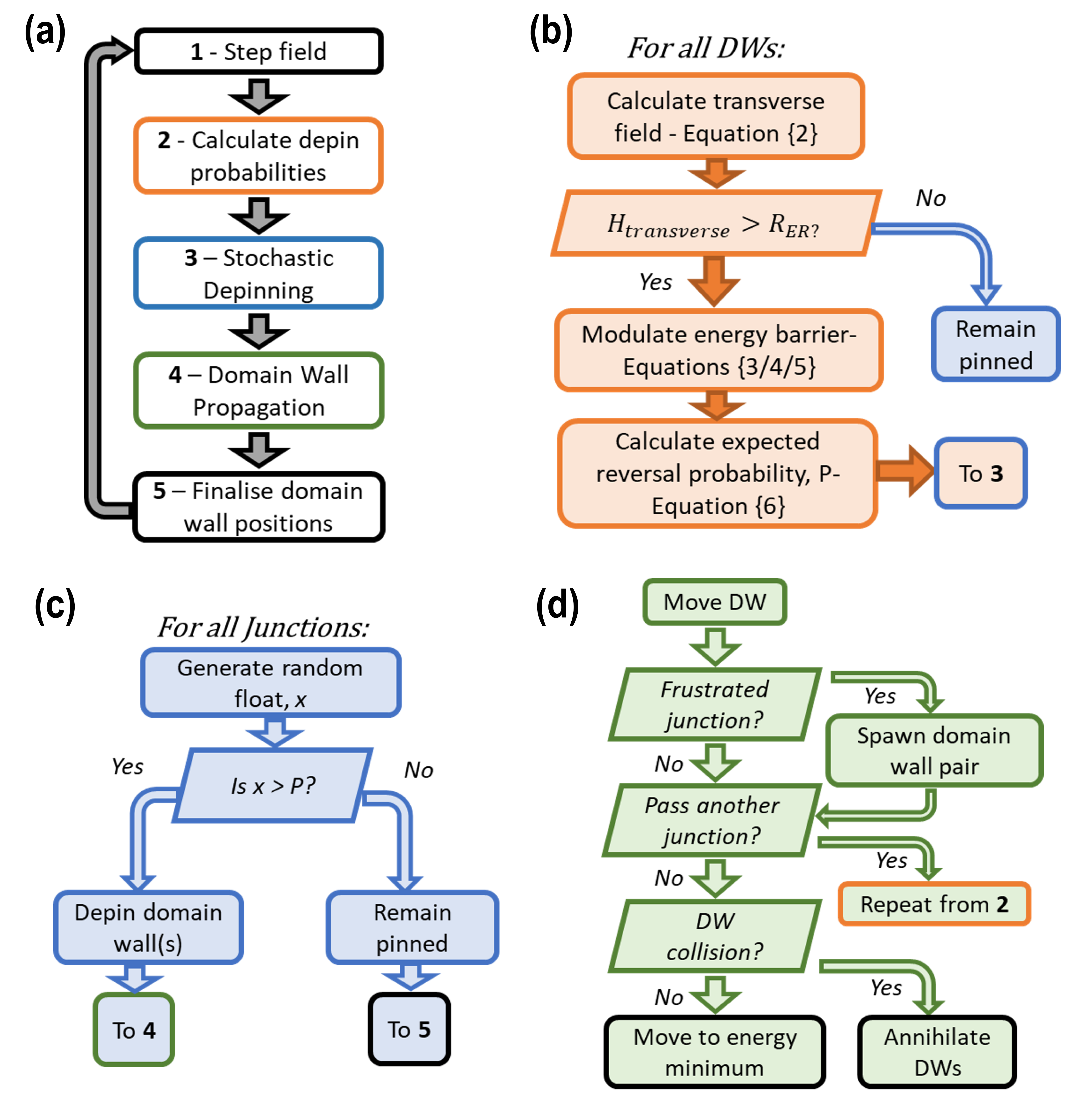}
    \caption{(a)- Overview of the key steps the taken in the modelling procedure. (b)-Process for determining the probabilities P for each of the domain walls to propagate within the model. (c)-Process for deciding the stochastic outcomes of pinning events based on previously calculated probability P. (d)- Process for determining the next state of the array from the outcome of depinning events, and any collisions or additional junctions that may be passed on the domain wall's path to the energy minimum.}
    \label{fig:fig3}
\end{figure}
Figure 3(a) shows how the model evolves the magnetic state of the system: Firstly, the external applied field is moved by a fixed angular step. The relative probabilities of all domain walls in the system depinning are then calculated via equations $(1)$ to $(6)$, and compared with a random variable drawn from a uniform distribution between 0 and 1 to determine which domain walls are free to propagate. The domain walls then propagate towards their respective energy minima along the field vector, potentially interacting with other domain walls and junctions as they move.\\ \\
Figure 3(b) shows a flowchart for the process of calculating depinning probabilities. In RingSim, two sources of domain wall pinning are considered: pinning due to edge roughness and pinning at junctions. The effects of edge roughness are included by imposing a threshold field, below which domain wall propagation does not occur \cite{dutta2017}. This information is stored within a vector representing the edge roughness threshold field for each ring, $R_\mathrm{ER}$, created by sampling from a normal distribution with a fixed mean and standard deviation to resemble the expected variance in properties via manufacturing imperfections in experimental samples. The exact mean and variance were determined via correlation with experimental data as discussed in section 5. This acts as a hard threshold by setting depinning probability to zero if $H_\mathrm{transverse}$ (via equation $(2)$) is below the value of $R_\mathrm{ER}^i$ for a given ring i. \\ \\
Domain wall pinning at junctions is modelled as follows: Domain walls occupying the same junction are considered as coupled in RingSim, with the calculation of reversal probabilities occurring only once with coupled outcomes for both domain walls. The number of domain walls at each junction is given by the entry in $J_\mathrm{DW}$ for a particular junction. Each junction has its own energy barrier and switching field, stored in the vectors $J_E$ and $J_H$, which are scaled for domain wall-domain wall interactions via equations $(4) \& (5)$ if two domain walls occupy the junction. The effects of the external field are accounted for by scaling the energy barrier via equation $(3)$, giving the magnitude of the effective energy barrier $\Delta E$. As the depinning process is thermally activated, the expected transition time for reversal is determined via equation $(1)$. The rotational frequency of the applied field is then used to give an expected probability of reversal for a given angular step, described via equation $(6)$. \\ \\
Figure 3(c) shows a flow chart for determining whether a domain wall depins during a given step of the model. A random number sampled from a uniform distribution between zero and one, $x$, is generated for each junction, and is compared to $P_\mathrm{depin}$. Domain walls occupying any junction where $P_\mathrm{depin} > x $ are deemed free to propagate. The process for domain wall propagation is outlined in figure 3(d). Depinned domain walls propagate either clockwise or anticlockwise according to the shortest route to their respective energy minima. If the propagation of a domain wall would lead to magnetic frustration across a junction, then the nucleation process occurs. A pair of domain walls, one head-to-head and one tail-to-tail,  are initialised at the junction in the adjacent ring. The domain wall closest to its respective energy minimum moves directly to this location. The other domain wall is then flagged for an additional depinning check at the junction where it was initialised. In the case where both minima are equidistant from the initialisation junction, both domain walls propagate to their respective minima where they remain until the next field step. \\ \\
In a given ring, domain walls propagate towards their energy minimum until one of three conditions are met: (a) If they collide with another domain wall before reaching their respective minimum, the domain walls annihilate, and leave the ring in a vortex state. (b) If the domain wall reaches another junction before its energy minimum, then the domain wall is flagged for an additional depinning check at that junction since it may become locally pinned again. (c) If the domain wall reaches its energy minimum, it remains there until the next field step. \\ \\
After propagation has occurred, the junctions are then checked for the number of domain walls occupying them, and the $J_{DW}$ vector is updated accordingly. The magnetisation state of the array is then calculated from the final positions, and returned in terms of normalised $M_x$ and $M_y$ components, and all state matrices are updated. The field then moves another angular increment, and the process starts over. \\ \\
\section{Fitting Model Parameters}
RingSim has a number of free parameters that must be appropriately selected to recreate experimental behaviours. Here, data is gathered from (and the model fitted to reflect) manufactured devices with nominal diameters of $4 \mathrm{\mu m}$, track widths of $400$ nm, thicknesses of $10$ nm, and with each ring overlapping $50{\%}$ of its track width with its neighbour.   Values of $\tau_0$ and $\alpha$ were taken from literature, as an attempt frequency of $1$ GHz for Permalloy \cite{lau2007common}, and an alpha value of $3/2$ \cite{wegrowe_magnetic_1997}. The remaining free parameters were fitted to reproduce experimental or other simulation data, with the process described here.\\ \\
To establish the effects of domain wall-domain wall interactions on the switching field of the junction in absence of thermal effects, micromagnetic simulations using the MuMax3 \cite{vansteenkiste_design_2014} software package were performed on a pair of overlapping half-rings, representing a single junction, but extending to ring properties via symmetry. Material parameters of the system were set to reflect $\mathrm{Ni}_{80}\mathrm{Fe}_{20}$ in line with the manufactured devices ($M_{s} = 860$ kA/m, $A_{ex} = 13$ pJ/m), with an artificially large damping parameter, $\alpha_{G} = 1$,  to speed up simulation convergence. The simulations, shown in Figures 4(a)$\And$(b), were initialised with a single domain wall and two domain walls occupying the junction respectively, and field tangential to the junction was ramped in $1$ Oe increments every 8 ns. The domain walls were deemed to depin at the field when they became fully delocalised from the junction, and the strength of the applied field recorded, with depinning fields of $79$ Oe and $64$ Oe observed for the one and two domain wall cases respectively. While these results reflect the zero-kelvin switching field for an idealised material, imperfections from the manufacturing process (lower true saturation magnetisation, imperfect geometry, presence of grains etc.) mean these values will not be numerically identical to those of a manufactured device. However, these values were used to be indicative of the ratio between the two processes, and hence used to determine the value of $k$ in equation $(4)$. For simplicity, the scaling of the energy barriers were assumed to be equivalent, giving a $k$ value of 0.81 for equations $(4)$ and $(5)$. \\ \\
\begin{figure}[ht!]
    \centering
    \includegraphics[width=\textwidth]{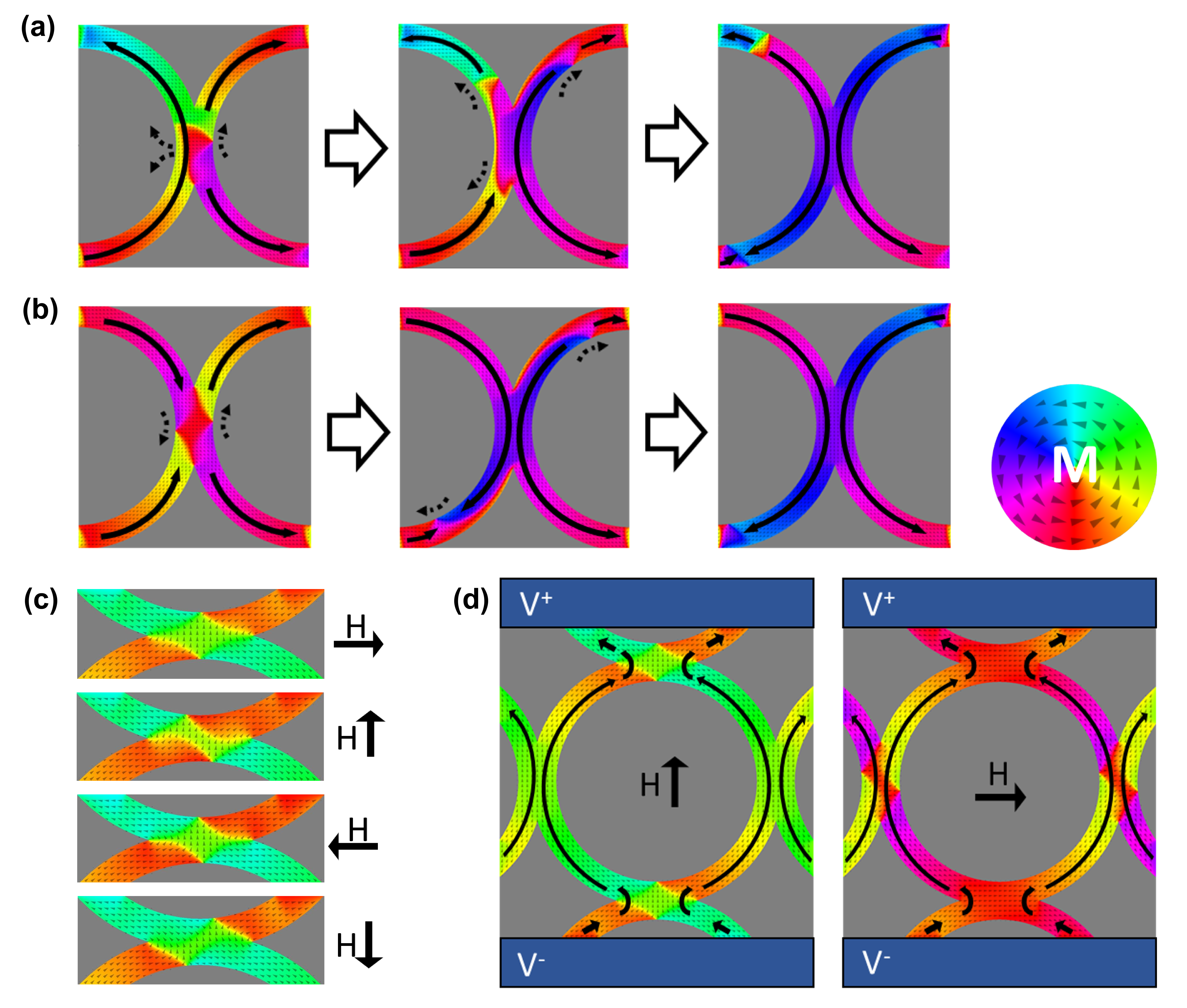}
    \caption{(a)/(b)- Domain wall reversal processes for one and two domain wall cases respectively, produced using MuMax3. Black arrows show direction of domains, white arrows reflect increasing of the applied field. (c)- Mumax simulations of domain wall structure across a junction during applications of 20 Oe rotating field in $\pi/2$ radian steps, showing the expansion and contraction of the domain wall via magnetic susceptibility. (d)- MuMax simulations of domain state with domains pinned in orientations parallel to current density (left) and orthogonal to current density (right). Black arrows reflect flow of current density. Colour wheel reflects direction of magnetisation in all plots, with local direction shown by grey triangles on the colour wheel.}
    \label{fig:fig4}
\end{figure}
The remaining free parameters of $R_{ER}$, $E_0$, and $H_{sw}^0$ were fit to magnetoresistance (AMR) measurements of the nanorings. As described in previous works \cite{vidamour2023reconfigurable}, the AMR response of the nanoring array has two distinct responses to rotating fields, one at the frequency of the rotating field (1f response) and another at twice the field frequency (2f response). The 1f response occurs due to susceptibility effects, with elastic stretching and contraction of the domain walls in the system in response to the rotating field (Figure (c)). The 2f signal depends upon the propagation of domain walls as they move between junction sites where pinned domain walls sit either orthogonal or parallel to the current density (Figure 4(d)). \\ \\
Figure 5(a) shows the relative magnitudes of the Fourier components of the AMR response of the 1f and 2f frequency components across a range of applied fields over 30 rotations. Two key features of this response were used to fit model parameters: Firstly, the end of the linear regime of the 1f response reflected the onset of domain wall motion, as the change from linear increase is due to the addition of incomplete propagation of domain walls around the rings. This allows determination of the mean value of $R_{ER}$ as $20.5$ Oe. Secondly, the magnitude of the 2f signal reflects the number of domain walls propagating in the system. This is proportional to the amplitude of the net magnetisation oscillations of the array for a given rotation, which was given by $M_x$ and $M_y$ in RingSim. This equivalence allowed tuning of the remaining free parameters of $E_0$ and $H_0$ by comparing the amplitude of the magnetisation output in RingSim to the experimental 2f data for the same applied fields, and selecting $E_0$ and $H_0$ values which provide the same response. The experimental procedure used to generate the 2f data in experiment (30 rotations of applied fields between 25-30 Oe) was simulated in RingSim, for a series of initialisations spanning a range of  $E_0$ and $H_0$ values. \\ \\
\begin{figure}[ht!]
    \centering
    \includegraphics[width=150mm]{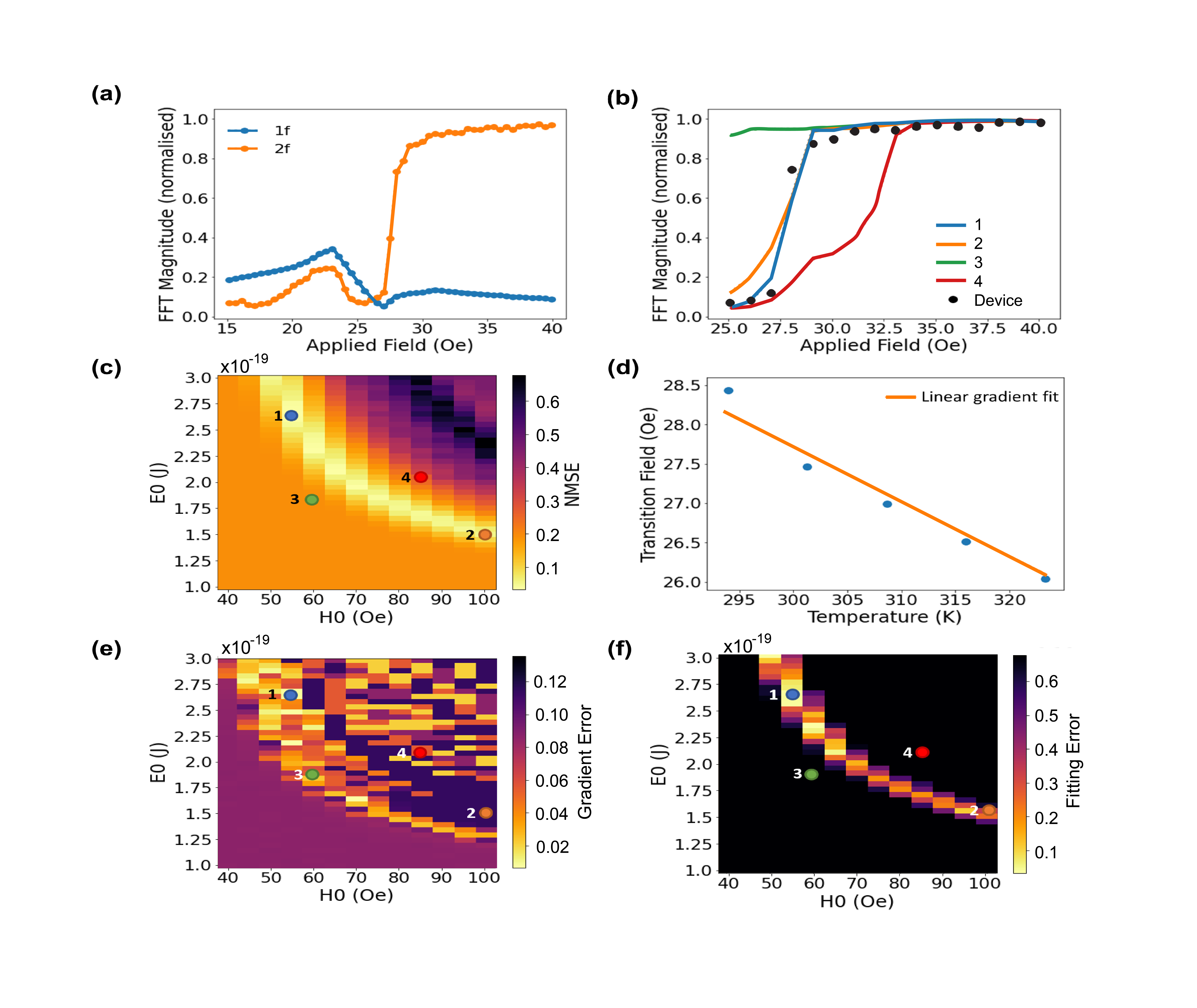}
    \caption{(a)- Normalised magnitudes of Fourier components corresponding to the clock frequency (1f) and twice the clock frequency (2f) over 30 rotations of varying magnetic field. (b)- Example fits of RingSim magnetisation output to the 2f transition in experimental device. The four coloured lines reflect different $E_0/H_0$ pairs, shown in the map in (c). (c)- Colourmap of mean-squared error between model's reproduction of magnetisation and experimentally gathered response via AMR for all explored $E_0/H_0$ values. (d)-Plot of extracted transition fields by varying the temperature of the magnetic ring array. Transition field determined from the point of maximum gradient in the 2f response, with linear fit used to extract the change in transition field with temperature shown. (e)-Colourmap showing gradient of transition field with respect to temperature, shown for all $E_0/H_0$ pairs. (f)- Combined fitting metric made by combining difference in gradient of transition field between simulated system and experimental system and the mean-squared error between the simulated 2f transition and experimental data.}
    \label{fig:fig5}
\end{figure}
Figures 5(b) and 5(c) show the magnitude of the oscillating $M_y$ signal from RingSim over a few example $E_0$ and $H_0$ pairs with respect to applied field compared with the device's 2f response, and the mean-squared error between the simulated magnetisation and the experimental response across all $E_0$ and $H_0$ pairs respectively. It can be observed that a band of $E_0$ and $H_0$ pairs are able to fit the experimental data well, reflected by the region of low mean-squared error in Figure 5(c). \\ \\
Since temperature also modulates the relative depinning probabilities which determine the number of propagating domain walls, the AMR response of the experimental system over a range of applied temperatures must also be determined to find the specific $E_0$ and $H_0$ pair that describes the system best.  The temperature of the system was was controlled by mounting the device on a Peltier cell, with temperature measured via a pyrometer positioned above the device. From these measurements, a linear shift in the point of maximum gradient of the 2f response, hereon termed the 'transition field',  was observed, and shown in Figure 5(d). Similarly to the previous fitting process, these experiments were repeated within RingSim, and the gradient of this linear shift calculated across a range of $E_0$ and $H_0$ pairs and compared to the experimentally gathered data. To determine the magnitude of the shift of characteristic transition points with respect to temperature, the point of maximum gradient in the 2f transition in experiments (or $M_y$ amplitude in RingSim) was chosen. Crucially, the error between simulated and experimentally measured gradient of the shift in transition field with increasing temperature shown in \ref{fig:fig5}(e) is different when compared to the 2f fit shown in \ref{fig:fig5}(c). This additional constraint enabled selection of a single $E_0$ and $H_0$ pair from the range of viable solutions in \ref{fig:fig5}(c)  An optimal $E_0$ and $H_0$ pair was chosen which reconciles both experiments, here determined to be $H_0 = 55 $ Oe and $E_0 = 2.625 \times 10^{-19} $ J (point '1' in Figure 5). 
\section{Validating the Model}
\begin{figure}
    \centering
    \includegraphics[width=140mm]{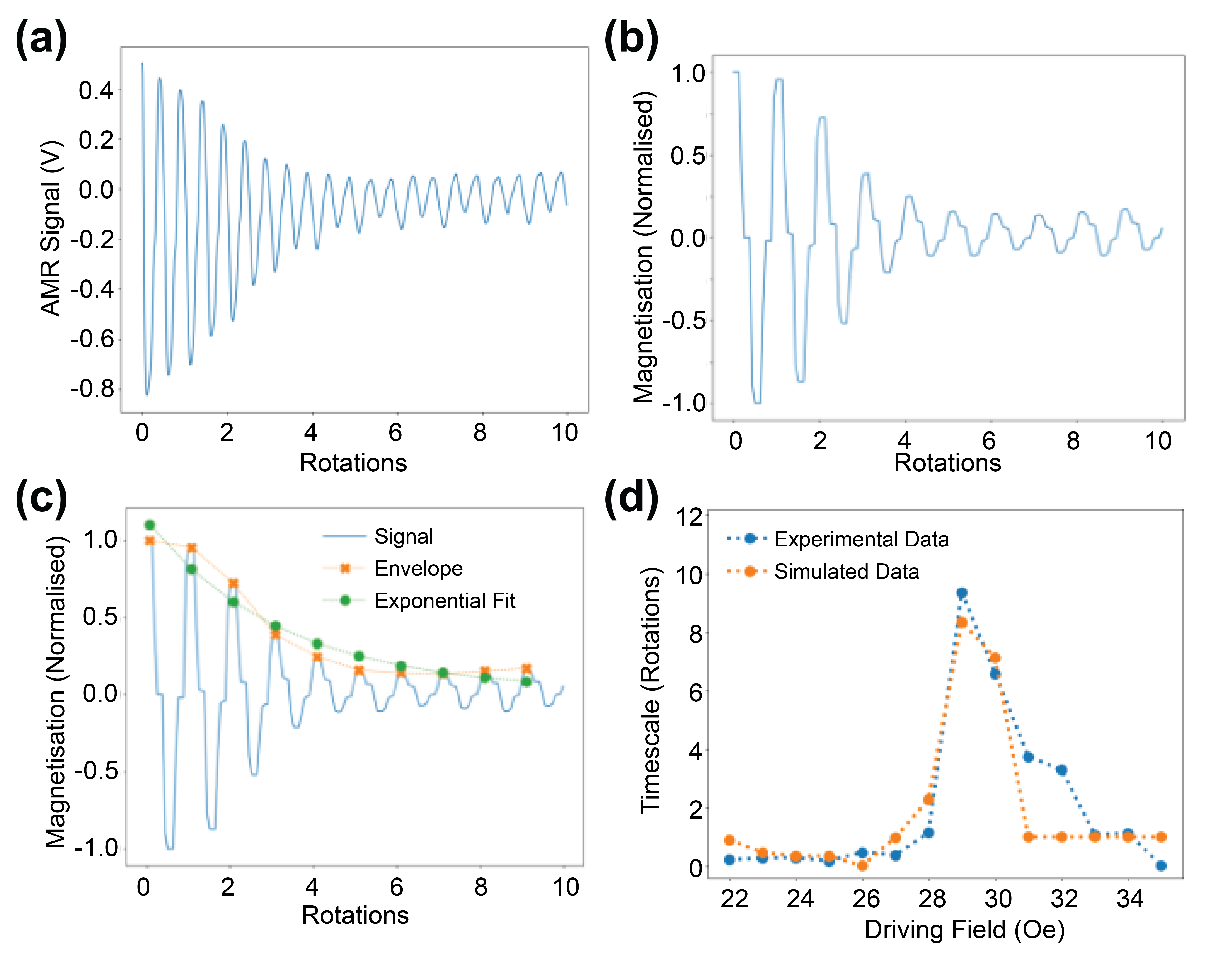}
    \caption{(a)- Measured voltage signal from experimental measurement of the ring devices under 10 rotations of 29 Oe applied field. (b)- Simulated magnetisation response via RingSim for 10 rotations of 28 Oe applied field. (c)- Outline of the procedure for measuring dynamic timescales. First, the envelope of the underlying signal is generated by marking the maximum magnetisation over a rotation of applied field, shown by the orange markers. Then, an exponential fit is generated to replicate the envelope of the magnetisation/AMR signal, shown in green. (d)- Comparison of the resulting decay timescales for experimentally gathered data (blue), and the simulated magnetisation output (orange). Timescale is presented with respect to number of rotations rather than in time.}
    \label{fig:fig6}
\end{figure}
The previous section outlined a procedure for fitting model parameters to the AMR response of the rings following 30 rotations of the applied field, where the DW population was expected to have reached equilibrium. However, the ring arrays are known to exhibit complex transient dynamics over several rotations of applied field, as well as different microstate populations of the three ring domain states (Figure 1(b)) with respect to driving field \cite{dawidek_dynamically-driven_2021}. Thus, to confirm that the model also captures these behaviours, the model's predictions were validated against additional experimental data that measure the dynamic timescales of the system's response, as well as the populations of domain states observed experimentally. \\ \\
To establish the dynamic behaviours of the physical system, further AMR measurements were performed to determine the rate at which the ring array reaches dynamic equilibrium in its AMR response from a saturated state over a range of applied fields. Figure 6(a)/(b) show the AMR response of the physical device at 29 Oe and the equivalent magnetisation response generated by RingSim respectively. Although the RingSim magnetisation response occurs at half the frequency of the AMR response, the signals are very similar in terms of decay time and magnitude of oscillation at equilibrium.  In order to evaluate these dynamic timescales $\tau_d$, the envelope of the AMR was calculated over successive rotations, and an exponential function of the form $X(t) = X_0 - a e^-\frac{t}{\tau_d} + c$ was fitted to the resulting decay curve. This was compared to a similar exponential fit to the envelope of the magnetisation signal generated by RingSim, shown in Figure 6(c).\\ \\
Figure 6(d) shows the comparison between the fitted $\tau_d$ parameters across these fields for both simulated and experimental data. There is excellent agreement between the simulation and experiments, showing that RingSim effectively simulates the regions of highly stochastic propagation well, and corroborating the presence of the longest timescale at H = 29 Oe. However, there are longer timescales observed H = 30-32 Oe in experiments than in simulation, with the simulation predicting the equilibrium amplitude is reached instantaneously. This suggests that RingSim underestimates the field at which domain walls deterministically overcome pinning and magnetisation oscillates at maximum amplitude. Further evidence of this can also be observed in 5(b) where the best-fit data from RingSim saturated more quickly than the experimental FFT magnitude.  \\ \\
To explore the microstates formed by the ring array, X-ray photo-emission electron microscopy (X-PEEM) was performed on subsections of the nanoring array at Beamline I06 at Diamond Light Source, UK. These measurements were performed by initialising the array to 'onion' state rings with a strong pulse of magnetic field, before driving with 30 rotations of applied fields at varying strengths. Images of the resulting domain state were generated by averaging a series of X-ray absorption (XAS) images on and off the Fe-$\rm{L}_{3}$ resonance with clockwise and anti-clockwise polarised X-rays, generating contrast according to magnetisation direction via X-ray magnetic circular dichroism  (XMCD). Figure 7(a) (upper) shows example X-PEEM micrographs of the arrays at various applied fields.\\\\
To validate these behaviours in RingSim, a visualisation tool was developed which emulates the magnetic contrast observed in X-PEEM according to the values of $M_y$ in each segment of all of the rings in the simulated system. The experimental procedure outlined for X-PEEM (saturation, 30 rotations of applied field) was repeated within RingSim, and compared with images generated via the visualisation tool. Figure 7(a) compares images measured by XPEEM with those generated by RingSim at five different applied fields. RingSim shows similar grouping of larger domains locally, reflecting a similar tendency for the domain wall-domain wall interactions in both the physical device and within RingSim to lead to local regions of magnetic order, suggesting that RingSim is able to encapsulate the emergent behaviours observed in experiments.  \\\\
To quantify the relative proportions of vortex, onion, and three-quarter state rings from the previous images, a custom image processing library within python\cite{PEEM-gitlab-repo}  was used to identify each of the ring states and count their populations. In RingSim, populations were determined from the number and locations of domain walls within the model. Figures 7(b) and 7(c) show the variation of these populations for simulated 25x25 ring arrays and experimentally gathered data on a sub-sample of the 25x25 array respectively. RingSim captures the general population trends for the ring states with respect to the applied field well. However, there is some variance between the relative populations observed in experiments and simulated by RingSim. \\ \\
One possible source of the differences between the X-PEEM measurements and RingSim model is that the sample used for X-PEEM is different to the sample used for fitting the model. Although the nominal dimensions of the arrays' designs were the same, they were manufactured in different lithography and deposition runs from the electrically contacted arrays, which could lead to some slight variation in ring width/thickness, accounting for the slight shifts in field. Additionally, remanence in the iron cores of the electromagnets used to generate the applied fields in the experimental data may have led to slightly asymmetrical field rotations, which could explain a biasing in the formation of vortex states in the experimental data which is not seen in the model. However, in combination with the previously presented results, there is strong evidence that RingSim provides an excellent description of the overall processes that dictate the response of the magnetic nanoring arrays.\\ \\
\begin{figure}[ht!]
    \centering
    \includegraphics[width=\textwidth]{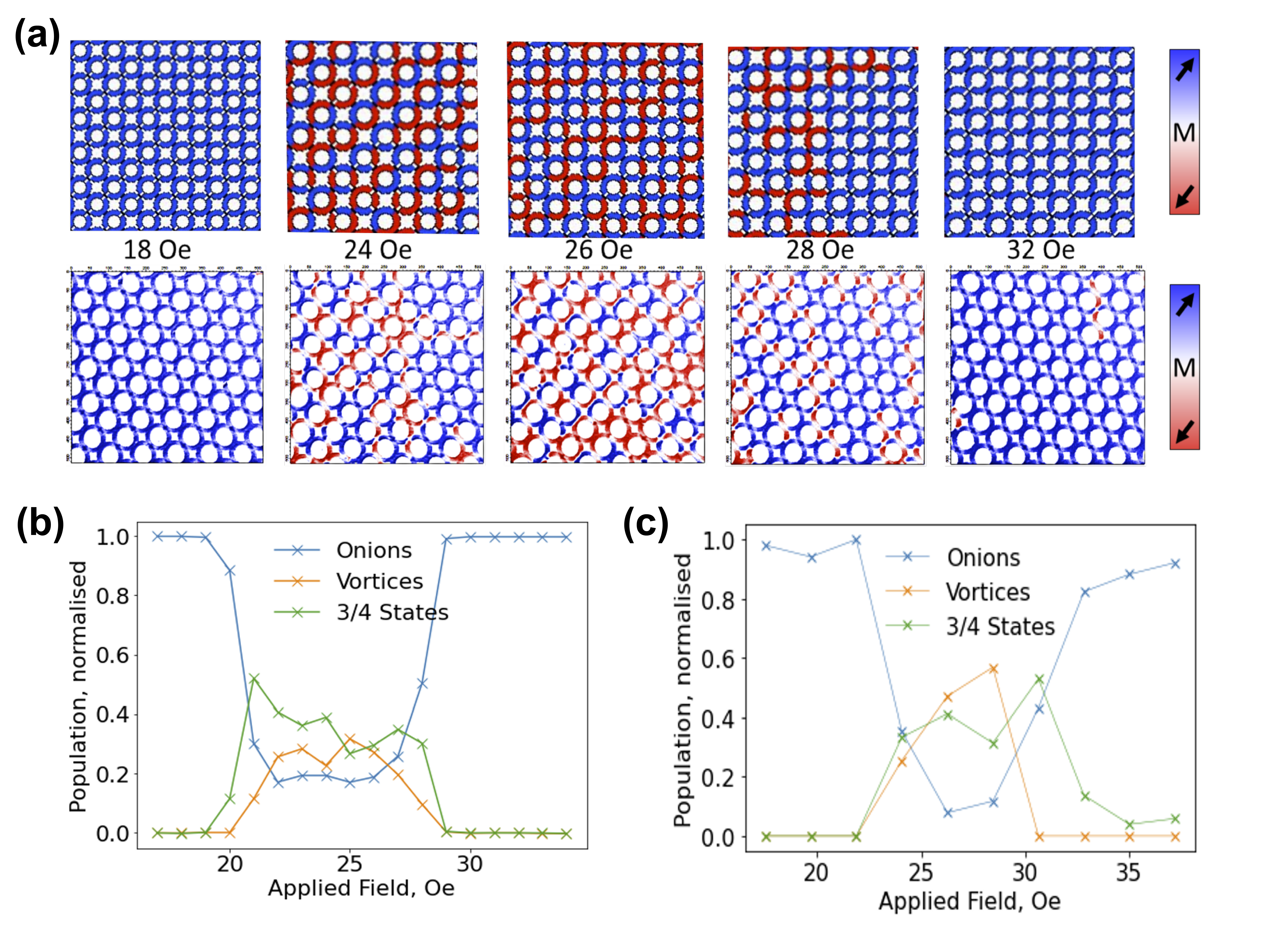}
    \caption{(a)- Comparison of magnetisation state between ring structures generated via RingSim (upper) and experimentally gathered data via X-PEEM imaging (lower) for applied fields between 18 and 32 Oe. In both cases, colour reflects magnetisation direction along the vertical axis, reflected in the colour bar on the right. (b)- State count of the three different ring domain configurations (onion, vortex, and three-quarter) with respect to the magnitude of the simulated rotating field. Generated over a 25x25 array, normalised against ring number. (c)-State count of ring domain configurations with respect to the magnitude of current provided to driving electromagnet \cite{foerster_custom_2016}. Generated over subsection of 25x25 ring array containing 40-50 nanorings, normalised against ring number for a given image.}
    \label{fig:Figure7}
\end{figure}
\section{Conclusions}
In this paper, we have outlined an agent-based methodology for modelling extended nanowire networks. With a combination of analytical models for calculating reversal probabilities and directly-programmed phenomenological behaviour, RingSim was able to emulate the behaviours observed in real nanoring devices.\\ \\
The resulting model provides excellent agreement with experimentally gathered data, not only in the global response of the system following equilibration of its DW population, but also with the dynamic timescales associated with reaching equilibrium as well as the typical domain microstates that are formed across different driving fields. RingSim is able to model relatively large areas of magnetic materials with modest computational overheads, allowing predictions to be made that would be practically impossible to achieve with general-purpose micromagnetic simulators. As a rough benchmark of performance, RingSim is able to simulate the response of a 25-by-25 array of nanorings at speeds of 1.4 rotations per second on an Intel i5 processor, allowing rapid evaluation of device response to arbitrary field inputs. \\ \\
While the exact formulation and phenomenology featured within RingSim is specific to the system of interconnected magnetic nanorings, we believe that the general methodology of reducing the simulated magnetic processes to the modelling key agents within the system, and programming interactions phenomenologically, can be applied to many other similar systems, such as connected artificial spin-ice networks, domain wall logic networks, racetrack memory etc. Thus, simulation tools such as RingSim can be useful for rapid exploration of mesoscale device behaviours that lie beyond the capabilities of conventional simulation approaches.
\section{Author Contributions}
ITV designed and programmed RingSim, and performed all simulations with RingSim. MuMax simulations were performed by ITV and GV. GV performed quantitative analysis on X-PEEM images. DG helped optimise the speed of the model. ITV, GV, and CS designed the setup for performing AMR measurements. ITV performed all AMR measurements. GV, CS, and PWF designed and manufactured all samples. ITV, RMRR, AW, TJH, and DAA performed X-PEEM measurements, which were overseen by DB, FM, and SSD. DAA and TJH conceptualised the work. 
\section{Acknowledgements}
The authors thank STFC for beam time on beamline I06 at the Diamond Light Source, and thank Jordi Prat, Michael Foerster and Lucia Aballe from ALBA for providing quadrupole sample holders \cite{foerster_custom_2016}. I.T.V. acknowledges DTA-funded PhD studentships from EPSRC. The authors gratefully acknowledge the support of EPSRC through grant EP/S009647/1, EP/V006339/1 and EP/V006029/1. This project has received funding from the European Union’s Horizon 2020 FET-Open program under grant agreement No 861618 (SpinEngine). 
\label{Bibliography}
\bibliography{bib.bib}
\end{document}